\documentclass[a4paper,fleqn,usenatbib]{mnras}

\usepackage{txfonts}

\usepackage[T1]{fontenc}
\usepackage{ae,aecompl}

\usepackage[toc,page]{appendix}
\usepackage{natbib}
\usepackage{amssymb}
\usepackage{epsf}
\usepackage{pstricks}
\usepackage{color}
\usepackage{graphicx}
\usepackage{slashed}
\usepackage{relsize}
\usepackage{morefloats}
\usepackage{multirow}
\usepackage{array}
\usepackage{hyperref}
\usepackage[justification=centering]{caption}
\usepackage{url}
\usepackage{subfig}
\title[Detecting Pulsars with Interstellar Scintillation in Variance Images]{Detecting Pulsars with Interstellar Scintillation in Variance Images}

\author[S. Dai et al.]{
S. Dai$^{1}$\thanks{E-mail: shi.dai@csiro.au},
S. Johnston$^{1}$,
M. E. Bell$^{1}$,
W. A. Coles$^{2}$,
G. Hobbs$^{1}$,
R. D. Ekers$^{1,3}$,
E. Lenc$^{4,5}$
\\
$^{1}$CSIRO Astronomy and Space Science, Australia Telescope National Facility, Box 76 Epping NSW 1710, Australia\\
$^{2}$Department of Electrical and Computer Engineering, University of California, San Diego, La Jolla, CA 92093, USA\\
$^{3}$International Centre for Radio Astronomy Research, Curtin University, Bentley, WA 6102, Australia\\
$^{4}$Sydney Institute for Astronomy, School of Physics, The University of Sydney, NSW 2006, Australia\\
$^{5}$ARC Centre of Excellence for All-sky Astrophysics (CAASTRO), Redfern, NSW, Australia\\
}

\date{Accepted XXX. Received YYY; in original form ZZZ}

\pubyear{2016}

\begin{document}
\label{firstpage}
\pagerange{\pageref{firstpage}--\pageref{lastpage}}
\maketitle

% Abstract of the paper
\begin{abstract}

Pulsars are the only cosmic radio sources known to be sufficiently compact to show diffractive 
interstellar scintillations.   Images of the variance of radio signals in both time and frequency 
can be used to detect pulsars in large-scale continuum surveys using the next 
generation of synthesis radio telescopes. This technique allows a search over the full field of view 
while avoiding the need for expensive pixel-by-pixel high time resolution searches.
We investigate the sensitivity of detecting pulsars in variance images. We show that variance 
images are most sensitive to pulsars whose scintillation time-scales and bandwidths are close 
to the subintegration time and channel bandwidth. 
Therefore, in order to maximise the detection of pulsars for a given radio continuum survey,  
it is essential to retain a high time and frequency resolution, allowing us 
to make variance images sensitive to pulsars with different scintillation properties.
We demonstrate the technique with Murchision Widefield Array data and show that variance images 
can indeed lead to the detection of pulsars by distinguishing them from other radio sources.

\end{abstract}

% Select between one and six entries from the list of approved keywords.
% Don't make up new ones.

%%%%%%%%%%%%%%%%%%%%%%%%%%%%%%%%%%%%%%%%%%%%%%%%%%

%%%%%%%%%%%%%%%%% BODY OF PAPER %%%%%%%%%%%%%%%%%%

\section{Introduction}

While pulsars are primarily detected and observed with high time resolution in order to 
resolve their narrow pulses, the phase-averaged emissions of many pulsars can be 
detected in previous radio continuum surveys~\citep[e.g.,][]{kca+98,ht99,k00}.
More importantly, continuum surveys are equally sensitive to all pulsars, not affected by the 
dispersion-measure (DM) smearing, scattering or orbital modulation of spin periods, and 
therefore allow us to search for extreme pulsars, such as sub-millisecond pulsars, 
pulsar-blackhole systems and pulsars in the Galactic centre.
A number of attempts have been made to search for pulsars in radio continuum surveys~\citep[e.g.,][]{kcc+00,ckb00}. 
Although the majority of these attempts have been unsuccessful, the 
first ever millisecond pulsar discovered, B1937+21, was initially identified in radio 
continuum images as an unusual compact source with a steep spectrum~\citep{bkh+82}. 

Next-generation radio continuum surveys, such as the ASKAP-EMU (Australian SKA Pathfinder-Evolutionary 
Map of the Universe)~\citep{nha+11}, LOFAR-MSSS (LOFAR-Multifrequency Snapshot Sky Survey)~\citep{hpo+15} 
and MWATS (Murchison Widefield Array Transients Survey)~\citep{bck+13}, will map a large sky area 
at different radio frequencies with high sensitivities (e.g., $\sim$10\,$\mu$Jy for EMU at $\sim1.4$\,GHz). 
Such surveys will necessarily detect a large number of pulsars in the images, and 
enable us to carry out follow-up observations and efficient targeted searches for the 
periodic signals. 
As we move towards the Square Kilometre Array (SKA) era, searching for pulsars in 
continuum images will complement the conventional pulsar search, and make it 
possible to find extreme objects.

The main challenge of detecting pulsars in continuum surveys or Stokes I images is to 
distinguish them from other unresolved point radio sources. Continuum surveys such as 
EMU will identify $\sim7\times10^{7}$ radio sources, while there are only $\sim1.2\times10^{5}$ potentially observable pulsars in our Galaxy~\citep[e.g.,][]{fk06}. Searching for pulsations 
from a large number of candidates will be very time-consuming, and therefore we need good 
criteria to select pulsar candidates. 
Although we know that pulsars have steep spectra and high fractions of linear and 
circular polarisation, these criteria are not exclusive as galaxies can also have steep spectra and 
be highly polarised. Also, as we average emission over the pulse phase, linear and circular 
polarisation of pulsars can be significantly lower in continuum surveys~\citep[e.g.,][]{dhm+15}.
However, pulsars are the only known sources compact enough to show diffractive interstellar 
scintillations (DISS), which distinguishes them from other radio sources. DISS are observed as 
strong modulations of pulsar intensities caused by the scattering in the ionised interstellar 
medium (IISM). The time-scales of DISS are of order of minutes and frequency scales are of order 
of MHz at $\sim1$\,GHz~\citep[e.g.,][]{r90}. Only recently have we had enough bandwidth and frequency 
resolution to detect DISS and next-generation continuum surveys will make it possible to search 
for pulsars as point sources showing strong intensity scintillations.

\citet{crd96} first suggested searching for pulsars in variance images and pointed out that the 
variance of pulsar signals can be introduced by pulse to pulse variability and both interplanetary 
and interstellar scintillations. However, they only focused on detecting pulsars with variance 
in time caused by pulse to pulse variabilities, which have variation time-scales of orders 
of milliseconds to seconds. In their work (using the Molonglo Observatory Synthesis Telescope 
at 843\,MHz with a bandwidth of 3\,MHz) they were unable to search for frequency variations 
because of their restricted bandwidth. In this paper, we will focus on the modulation of pulsar 
intensity in both time and frequency caused by DISS, and investigate detecting pulsar with 
DISS in variance images.   
In Section~\ref{diss}, we briefly review the basics of pulsar scintillation.
In Section~\ref{statistics}, we generally investigate the false alarm and detection 
probabilities and the detection sensitivity as a function of scintillation time-scales 
and bandwidths with simulations.
In Section~\ref{demo}, we demonstrate the technique with data taken with MWA.
We discuss our results and conclude in Section~\ref{discussion}.

%%%%%%%%%%%%%%%%%%%%%%%%%%%%%%%%%%%%%%%%%%%%%%%%%%%%%%%%%%%%%%%%%%%%%%%%%%%%

\section{Basics of pulsar interstellar scintillation}
\label{diss}

\begin{figure}
\center
\includegraphics[width=2.5in,angle=-90]{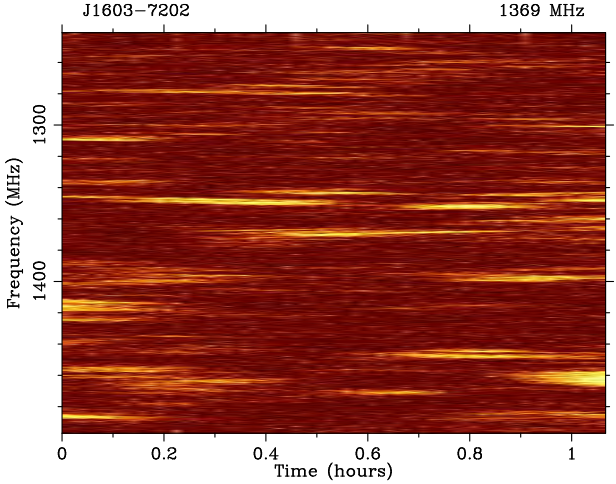}
\caption{Dynamic spectrum of PSR J1603$-$7202 observed with the Parkes telescope 
on 2009 August 12.}
\label{dynSpec}
\end{figure}

Pulsar signals are scattered as they propagate through the IISM
because of fluctuations in the electron density. One consequence of the scattering is  
the modulation of pulsar intensity as a function of time and frequency, which is called scintillation 
and can be observed in the dynamic spectrum. 
An example of the dynamic spectrum of PSR J1603$-$7202 is shown in Fig.~\ref{dynSpec}. The 
data was collected on the 2009 August 12 as a part of the Parkes Pulsar Timing Array 
project~\citep{mhb+13} and the data file can be obtained from the CSIRO data archive\footnote{\url{http://doi.org/10.4225/08/521616837BE48}}.
The dynamic spectrum is made using the PSRCHIVE software package~\citep{hvm04}. 
PSR J1603$-$7202 has a scintillation bandwidth of 5\,MHz and a time-scale of 582\,s
at a reference frequency of 1.4\,GHz~\citep{kcs+13}. In Fig.~\ref{dynSpec} we 
can see the strong modulation of pulsar intensity especially in frequency, which 
will result in a significant detection of this pulsar in variance images.
The theory of interstellar scattering and related observations have been reviewed by \citet{r90} and 
\citet{n92}.

Intensity scintillation is an inherently spatial process which is usually observed as 
time variation because the line of sight from the pulsar to the observer is moving through the 
spatial pattern of fluctuations in the electron density. In the regime of weak scintillation where the root-mean-squared fractional intensity 
fluctuation $m<1$, the spatial scale is $r_{\rm{f}}\approx \sqrt{L/k}$ and the time 
variation is similar throughout the observing band. Here $k=2\pi/\lambda$ is the wavenumber 
and $L$ is the distance from the scattering screen to the observer. As the scattering gets stronger $m$ 
increases, overshoots unity, and slowly relaxes back to unity when the scattering becomes very 
strong. In the very strong regime there are two spatial scales, $s_{\rm{dif}}$ and $s_{\rm{ref}}$, and, 
of course, the corresponding time scales. The two scales are related by $s_{\rm{dif}}\cdot s_{\rm{ref}}=r_{\rm{f}}^{2}$ 
and their separation increases as the strength of scattering increases. The DISS process 
becomes narrower in bandwidth $\delta\nu_{\rm{DISS}}$ as the scattering becomes stronger. The 
bandwidth is related to the spatial scales by $\nu_0/\delta\nu_{\rm{DISS}}=s_{\rm{ref}}/s_{\rm{dif}}$ 
where $\nu_0$ is the mean observing frequency. The refractive process (RISS) is relatively 
broad band. For typical pulsar observations the DISS can be seen in a dynamic spectrum (for example, Figure~\ref{dynSpec}), 
but the RISS is observed as day-to-day variations of the DISS. The observing bandwidth is 
normally much smaller than the central frequency (for instance, in Figure~\ref{dynSpec}, the bandwidth is 256\,MHz centred at a frequency of 1369\,MHz), and therefore $\delta\nu_{\rm{DISS}}$ 
and $\tau_{\rm{DISS}}$ do not show significant change across the band.

We consider a pulsar at distance $D$ from the Earth.  If we assume a thin scattering disc at 
$D/2$, an effective velocity of $V_{\rm{eff}}$ and a Kolmogorov spectra 
of the electron density fluctuations, the scintillation time-scale and bandwidth of DISS 
observed at a reference frequency $\nu$ can be estimated as~\citep[e.g.,][]{r77,gn85,cr98}
\begin{equation}
\tau_{\rm{DISS}}\propto\nu^{6/5}D^{-3/5}V_{\rm{eff}}^{-1},
\end{equation}
\begin{equation}
\delta\nu_{\rm{DISS}}\propto\nu^{22/5}D^{-11/5}.
\end{equation}
Therefore, the DISS time-scale and bandwidth increases as the observing frequency 
increases and as the pulsar distance decreases, but the DISS bandwidth changes much faster 
with the observing frequency and pulsar distance. For low frequency surveys, such as with the MWA and 
LOFAR, most pulsars will be in relatively strong scintillation, i.e., the scintillation 
time-scale and bandwidth will be much smaller than the integration time and observing bandwidth. 
In order to detect strong scintillations, we will need sufficient time and frequency resolution 
to resolve the scintillation, and as DISS bandwidth varies rapidly with observing frequency and 
pulsar distance we will normally need much higher frequency resolution than time resolution. 
For surveys at higher frequency, such as with ASKAP, while some pulsars will be in 
relatively weak scintillation, most distant pulsars will be in relatively strong scintillation, 
and therefore we will need enough bandwidth and integration time to cover the frequency 
and time-scales of scintillation, and also sufficient time and frequency resolution to 
detect strong scintillation pulsars. 
It is important that the subintegration time is < $\tau_{\rm{DISS}}$ and the channel bandwidth 
< $\delta\nu_{\rm{DISS}}$, otherwise the scintillation variance will be averaged out. Typically 
the time scale is not a problem but it is common for the channel bandwidth to limit the detection 
of scintillation. Scintillation of the most distant pulsars will not be observable 
because the channel bandwidth is too large.

In addition to the interstellar scintillation we note that interplanetary scintillation (IPS) and 
ionospheric scintillation may also be present in low-frequency observations ($<1$\,GHz). However, 
IPS will normally be in the weak scattering regime (even at low frequencies) unless the line-of-sight 
to the source is extremely close to the Sun. Ionospheric scintillation will only be in the strong 
scattering regime under active geomagnetic conditions. The scintillation bandwidth of both 
IPS and ionospheric scintillation will be much broader than the observing bandwidth used by 
telescopes such as MWA, and we therefore only expect to observe strong intensity variations 
in frequency caused by interstellar scintillation. The scintillation time-scale of IPS is typically 
shorter than one second~\citep[e.g.,][]{coles95}, while the scintillation time-scale of ionospheric 
scintillation is of orders of tens of seconds~\citep[e.g., 10 to 100\,s at 154\,MHz for MWA][]{lmb+15}. 
These time-scales are much shorter than that of DISS and the intensity variation will be averaged out 
by time integrations longer than a few minutes. Therefore DISS can be separated from IPS and ionospheric 
scintillation by their different time-scales and bandwidths.

\section{Detecting pulsars in variance images}
\label{statistics}

To investigate the detection of pulsars in variance images, it is important to understand the 
false alarm probability and the detection probability. In this section, through simulations and 
discussions of false alarm and detection probabilities, we aim to answer questions such as
\begin{itemize}
\item For a given survey with fixed total bandwidth, integration time, number of channels and subintegrations, 
how sensitive is it to pulsars with different scintillation time-scale and bandwidth?
\item For a given survey with fixed total bandwidth and integration time, how does the sensitivity change 
with different numbers of channels and subintegrations?
\end{itemize}

In our simulations and discussions below, we consider a radio continuum image of a sky area containing 
a pulsar. The total integration time is $T$ and the total bandwidth is $B$. Each image pixel 
consists of $N_{\rm{t}}$ subintegrations and $N_{\rm{f}}$ channels, and therefore the subintegration 
time is $\delta t=T/N_{\rm{t}}$ and the channel bandwidth is $\delta\nu=B/N_{\rm{f}}$. The standard 
deviation of noise averaged over $T$ and $B$ is $\sigma_{\rm{n}}$, which gives the standard deviation 
of noise in each dynamic spectrum pixel of $\sigma_{\rm{dyn}}=\sigma_{\rm{n}}\sqrt{N}$ and 
$N=\sqrt{N_{\rm{f}}N_{\rm{t}}}$. 
The pulsar shows DISS with a scintillation time-scale of $\tau_{\rm{DISS}}$ and a scintillation 
bandwidth of $\delta\nu_{\rm{DISS}}$. In the dynamic spectrum the pulsar flux density as a function 
of time and frequency is $S_{\rm{dyn}}(t,\nu)$, and the mean of $S_{\rm{dyn}}(t,\nu)$ over 
the dynamic spectrum is $S_{\rm{psr}}$. All parameters are in arbitrary units.

To simulate the dynamic spectrum of the DISS we assume relatively strong scattering. In this 
case the electric field of the DISS is a complex Gaussian process where the real and imaginary 
parts are uncorrelated. Thus the auto-covariance (ACF) of the intensity is the square of 
the auto-covariance of the field. There is a simple and exact analytical expression for the 
temporal covariance but only an approximate expression for the frequency covariance~\citep[e.g.,][]{crg+10}. 
We have grafted those expressions together in a way that preserves the existence condition 
for an ACF, that its Fourier transform be positive semi-definite,
\begin{equation}
C(t,\nu)=\exp{\left(-\frac{1}{2}\left[(\frac{t}{\tau_{\rm{DISS}}})^{\frac{5}{2}}+(\frac{\nu}{\delta\nu_{\rm{DISS}}})^{\frac{3}{2}}\right]^{\frac{2}{3}}\right)}.
\label{acf}
\end{equation}
The ACF of intensity is the square of this, so the $1/\mathrm{e}$ time and frequency scales are 
$\tau_{\rm{DISS}}$ and $\delta\nu_{\rm{DISS}}$ respectively. 

The Fourier transform of a Gaussian random process is also a Gaussian random process, so the 
Fourier transform of the electric field must be a Gaussian random process for which the 
expected value of the squared magnitude is the power spectrum $P$. We obtain $P$ from Fourier 
transformation of Eq.~\ref{acf}. We then create a realization of the electric field of the 
form $\sqrt{P/2}\cdot(a+\mathrm{i}b)$, where a and b are uncorrelated unit variance Gaussian random variables. 
The inverse transformation of this realization is the dynamic spectrum of the electric field 
$E(t,\nu)$, and its squared magnitude is a realization of the dynamic spectrum of intensity 
$I(t,\nu)=|E(t,\nu)|^{2}$.

This process provides a good approximation for most pulsars observed at centimetre or metre 
wavelengths. However it does not include the effects of enhanced refraction which are seen 
to cause correlation between the time and frequency variations in the dynamic spectrum. A 
more realistic simulation could be obtained with a full electromagnetic simulation~\citep{crg+10}, 
but such simulations would be difficult to extend to the very strong scintillation often 
seen at metre wavelengths.

\subsection{Noise, false alarms and a ``matched filter''}
\label{matchedfilter}

A detection is made when the detection statistic exceeds a threshold. The threshold 
is set so the probability that it is exceeded by noise alone is adequately small 
(of order per cent). Here we assume that noise is radiometer noise which is uncorrelated 
over the dynamic spectrum. The mean pulsar flux averaged over the dynamic spectrum $S_{\rm{psr}}$ 
gives a suitable detection statistic for continuum images or Stokes I images. The false 
alarm threshold would be a multiple of the standard deviation of the radiometer noise $\sigma_n$ 
averaged over the entire dynamic spectrum, i.e. detection is claimed if 
$S_{\rm{psr}}>C\cdot\sigma_{\rm{n}}$.

For a scintillating pulsar in variance images, detection would be claimed if 
\begin{equation}
V_{\rm{psr}}=S_{\rm{psr}}^{2}>C\cdot\rm{Std}(\sigma_{\rm{dyn}}^{2}),
\label{varDetection}
\end{equation} 
where $V_{\rm{psr}}$ is the mean of the sample variance of $S_{\rm{dyn}}(t,\nu)$ and 
equals to $S_{\rm{psr}}^{2}$ as $S_{\rm{dyn}}(t,\nu)$ follows an exponential distribution.
If there are $N\gg1$ independent samples in the dynamic spectrum and the radiometer noise 
is approximately Gaussian, the sample variance of $\sigma_{\rm{dyn}}^{2}$ computed over 
$N$ samples equals to $2N\cdot\sigma_{\rm{n}}^{4}$, and $\rm{Std}(\sigma_{\rm{dyn}}^{2})=\sigma_{\rm{n}}^{2}\sqrt{2N}$.
Therefore, Eq.~\ref{varDetection} gives a minimum detectable flux in a variance image of
\begin{equation}
S_{\rm{psr}}>\sqrt{C}\cdot\sigma_{\rm{n}}\cdot(2N)^{1/4}.
\end{equation}

The ratio $R$ of the minimum detectable flux in a variance image to the minimum detectable 
flux in a Stokes I image is $R=(2N)^{0.25}/C^{0.5}$. So in forming a dynamic spectrum 
and calculating the variance of the flux, one should not use more samples than necessary. 
On the other hand, if one uses fewer samples than there are independent fluctuations (scintles) 
$N_{\rm{indep}}$ in the scintillating flux, then the variance of the flux will be reduced by 
a factor of $\sim N/N_{\rm{indep}}$ smoothing. 

\begin{figure}
\center
\includegraphics[width=3in]{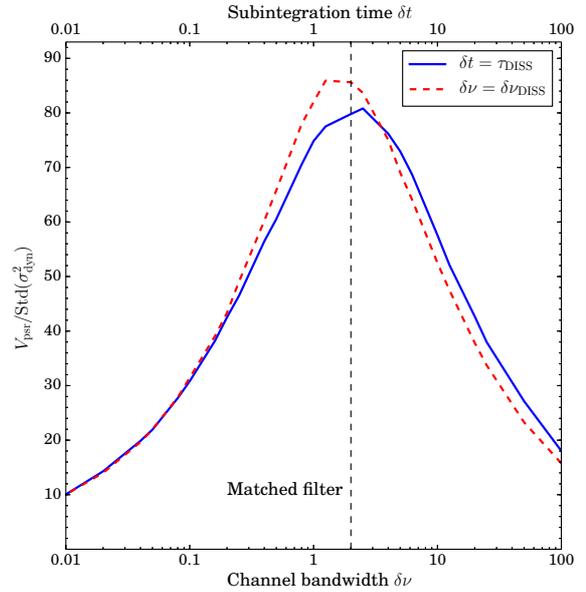}
\caption{$V_{\rm{psr}}/\rm{Std}(\sigma_{\rm{dyn}}^{2})$
as a function of channel bandwidth and subintegration time for a given pulsar 
in the variance image. The scintillation time-scale and bandwidth of the 
pulsar are $\tau_{\rm{DISS}}=1$ and $\delta\nu_{\rm{DISS}}=1$, respectively. The total 
integration time and bandwidth are $T=100$ and $B=100$, respectively. The noise level is 
$\sigma_{\rm{n}}=0.1$ and pulsar apparent flux density is $S_{\rm{psr}}=10$. For the blue 
solid line we set $\delta t/2=\tau_{\rm{DISS}}=1$ and vary channel bandwidth, while for the 
red dashed line we set $\delta\nu/2=\nu_{\rm{DISS}}=1$ and vary subintegration time.}
\label{matched}
\end{figure}

Clearly the optimal number of samples in the dynamic spectrum must be matched to the number 
of independent scintles in the scintillating flux. 
The time-scale and bandwidth of the DISS are auto-covariance widths. They should match 
the auto-covariance width of the sub-integration $\sim\delta t/2$ and the channel 
bandwidth $\sim\delta\nu/2$.
To demonstrate the ``matched filter'', we carry out simulations to show how 
$V_{\rm{psr}}/\rm{Std}(\sigma_{\rm{dyn}}^{2})$
varies with channel bandwidth and subintegration time. We set $T=100$, $B=100$, 
$\tau_{\rm{DISS}}=1$, $\nu_{\rm{DISS}}=1$, $S_{\rm{psr}}=10$ and $\sigma_{\rm{n}}=0.1$. 
In Fig.~\ref{matched}, for the blue solid line, we set $\delta t/2=\tau_{\rm{DISS}}=1$, 
which represents ``matched filter'' in time and gives a number of subintegrations of 
50. We simulate a dynamic spectrum with 50 subintegrations and 10000 channels, 
and then average over different numbers of channels to produce dynamic spectra with 
different channel bandwidths. We simulate 10000 realisations and calculate the 
mean of sample variance for each channel bandwidth. We separately simulate the 
noise following the same procedure and calculate the standard deviation of sample 
variance for each channel bandwidth. For the red dashed line, we set $\delta\nu/2=\nu_{\rm{DISS}}=1$
and carry out similar simulations for different subintegration time.
We can see in Fig.~\ref{matched} that both solid and dashed lines peaks at $\sim2$, 
which corresponds to the ``matched filter'' case that $\delta\nu/2\approx\delta\nu_{\rm{DISS}}$ 
and $\delta t/2\approx\tau_{\rm{DISS}}$.  

\subsection{Detection probability}
\label{detectionProb}

The detection probability is the probability of a signal exceeding the detection 
threshold. Different from that of a continuum source, the detection probability 
of a scintillating source will depend on its scintillation time-scale and bandwidth, 
and also on the integration time, bandwidth and time and frequency resolution of 
the survey. 
A useful way of understanding and investigating the detection probability 
is comparing the signal with its standard deviation. We can define the S/N of 
detecting a pulsar in Stokes I images as 
\begin{equation}
(\mathrm{S/N})_{\mathrm{I}}=S_{\mathrm{psr}}/\mathrm{Std}(S_{\mathrm{psr}}),
\end{equation} 
and the S/N of detecting a pulsar in variance images as 
\begin{equation}
(\mathrm{S/N})_{\mathrm{var}}=V_{\mathrm{psr}}/\mathrm{Std}(V_{\mathrm{psr}}).
\end{equation}
Assuming that there are $N\gg1$ independent samples in the dynamic spectrum and the 
noise is negligible, flux densities of a scintillating pulsar follow an exponential 
distribution, and we have 
\begin{equation}
\mathrm{Std}(S_{\rm{psr}})=S_{\rm{psr}}/\sqrt{N}, 
\end{equation}
\begin{equation}
\mathrm{Std}(V_{\rm{psr}})=\sqrt{\frac{(N-1)[(N-1)\mu_{4}-(N-3)\mu^{2}_{2}]}{N^{3}}}\approx\sqrt{8}\cdot S_{\rm{psr}}^{2}/\sqrt{N},
\end{equation}
where $\mu_{4}=9S_{\rm{psr}}^{4}$ is the fourth central moment and $\mu_{2}=S_{\rm{psr}}^{2}$
is the second central moment.
Therefore, (S/N)$_{\rm{I}}=\sqrt{N}$ and (S/N)$_{\rm{var}}=\sqrt{N/8}$, 
the detection possibility of a pulsar in variance images is a factor of 
$\sqrt{8}$ lower than that in Stokes I images. The underlying reason is that the 
variance of sample variance is larger than the variance of sample mean, and 
therefore the detection is more uncertain.

\begin{figure}
\center
\includegraphics[width=3in]{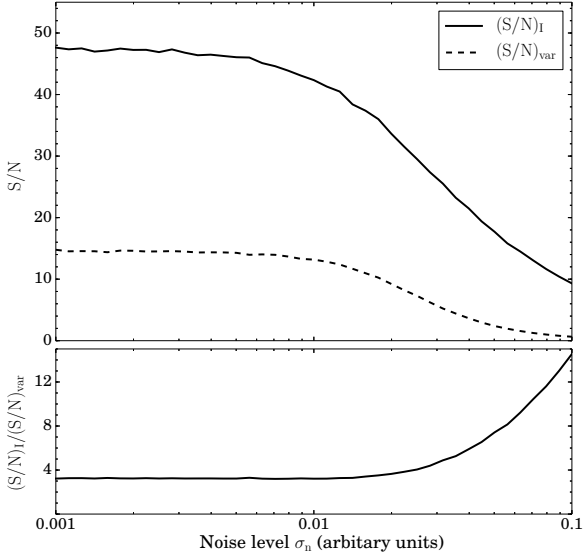}
\caption{$(\rm{S/N})_{\rm{I}}$ and $(\rm{S/N})_{\rm{var}}$ as a function of noise level in 
the upper panel and the ratio between them in the bottom panel. Parameters of the 
simulation are $T=100$, $B=100$, $\delta t/2=\tau_{\rm{DISS}}=1$, $\delta\nu/2=\nu_{\rm{DISS}}=1$ 
and $S_{\rm{psr}}=1$. }
\label{varStatistics}
\end{figure}

If the noise is not negligible, we find that the S/N of detecting pulsars 
in variance images drops faster than it does in Stokes I images as the noise level increases. 
In Fig.~\ref{varStatistics}, we show how (S/N)$_{\rm{I}}$ and (S/N)$_{\rm{var}}$ vary 
with the noise level in the upper panel and the ratio between them in the bottom panel. We set $T=100$, $B=100$, 
$\delta t/2=\tau_{\rm{DISS}}=1$, $\delta\nu/2=\delta\nu_{\rm{DISS}}=1$ and $S_{\rm{psr}}=1$. 
For each noise level in the figure, we simulate a dynamic spectrum of the pulsar 
with noises and calculate the sample mean and the sample variance. We assume 
that we have ten image pixels to measure the noise level, and therefore we 
separately simulate dynamic spectra with only noise for each of these ten pixel. 
We calculate the mean of sample mean and the mean of sample variance for 
these ten pixels and subtract them from the pulsar signal to remove noises. For 
each noise level in the figure, we simulate $10000$ realisations of above simulations 
and then calculate $\mathrm{Std}(S_{\rm{psr}})$ and $\mathrm{Std}(V_{\rm{psr}})$.
In Fig.~\ref{varStatistics} we can see that (S/N)$_{\rm{I}}$ is about a factor of 
$\sqrt{8}$ higher than (S/N)$_{\rm{var}}$ when the noise is negligible. As the noise 
becomes significant and increases, (S/N)$_{\rm{I}}$/(S/N)$_{\rm{var}}$ drops rapidly, 
which means that the variance image becomes less and less sensitive than the Stokes 
I image as the noise increases. 

We note that despite Stokes I images having higher sensitivity for detecting a pulsar, 
such images only provide limited information, e.g., the compactness, by which you can distinguish 
a pulsar from other radio sources. On the contrary, the DISS variance images are likely 
to allow exclusive detections of pulsars.

\subsection{Detection sensitivity of pulsars in variance images}
\label{detection}

To define the detection of a pulsar in variance images, we first determine the detection 
threshold as the value that is exceeded in only five per cent of the noise, which corresponds 
to a five per cent false alarm probability. Then we determine the flux density of a pulsar with which 
80 per cent of the measurements exceed the detection threshold. We define such a flux density as the 
sensitivity of a survey with a five per cent false alarm probability and 80 per cent detection probability.
Same detection statistics can also be defined for Stokes I images.

\begin{figure}
\center
\includegraphics[width=3in]{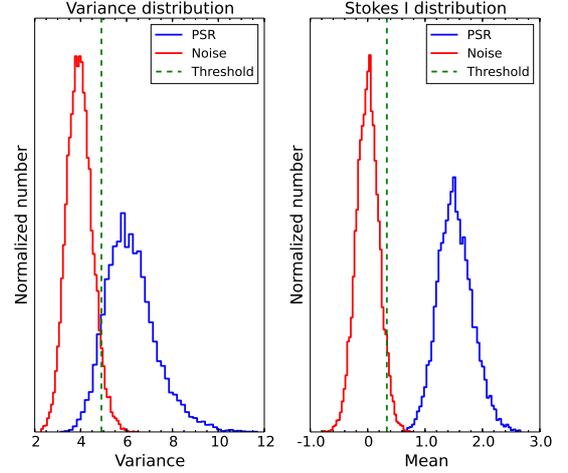}
\caption{Histograms of noise and pulsar intensities. The green dashed line shows the detection 
threshold corresponding to five per cent false alarm probability. Parameters of the simulations are 
$T=100$, $B=100$, $N_{\rm{t}}=10$, $N_{\rm{f}}=10$. $\tau_{\rm{DISS}}=1$ and $\delta\nu_{\rm{DISS}}=1$. 
The pulse flux density is set to be $S_{\rm{psr}}=1.5$ and the noise level is $\sigma_{\rm{n}}=0.2$.}
\label{histogram}
\end{figure}

We simulate dynamic spectra with noises to obtain the distributions of both mean flux 
density and variance of flux density of pulsars, and we separately simulate the noise distributions 
without pulsar signals. 
In Fig.~\ref{histogram}, we show an example of the distributions with 10000 simulations.  
We set $T=100$, $B=100$, $N_{\rm{t}}=10$, $N_{\rm{f}}=10$, $\tau_{\rm{DISS}}=1$ and $\delta\nu_{\rm{DISS}}=1$. 
The pulse flux density is set to be $S_{\rm{psr}}=1.5$ and the noise level is $\sigma_{\rm{n}}=0.2$.
The left panel shows results of variance images and the right panel shows results of Stokes I images. 
The green dashed line represents a detection threshold corresponding to five per cent false alarm 
probability. From the distributions we can see that, for the same false alarm probability, Stokes I 
images have a higher detection probability compared with variance images, which is consistent with our 
discussions in Section~\ref{matchedfilter} and \ref{detectionProb}. 

\begin{figure}
\center
\includegraphics[width=3in]{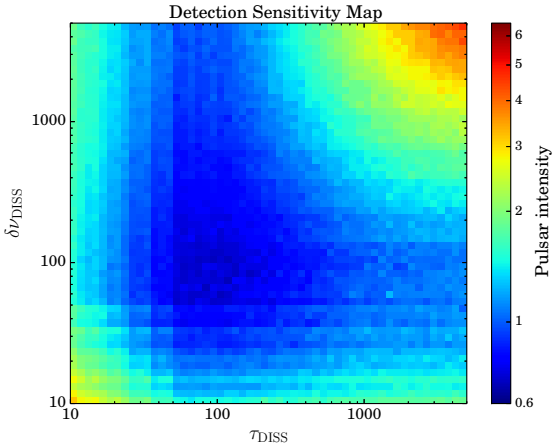}
\caption{Sensitivity map as a function of scintillation time-scale $\tau_{\rm{DISS}}$ and 
bandwidth $\delta\nu_{\rm{DISS}}$. Sensitivity are shown as the intensity in arbitrary units corresponding to 
a detection probability of 80\% and a false alarm probability of 5\%. Parameters of the simulation 
are $T=1000$, $B=1000$, $N_{\rm{t}}=10$, $N_{\rm{f}}=10$ and $\sigma_{\rm{n}}=0.1$. The sensitivity 
of a Stokes image with the same parameters is $\sim0.25$.}
\label{sensitivity}
\end{figure}

\begin{figure}
\center
\includegraphics[width=3in]{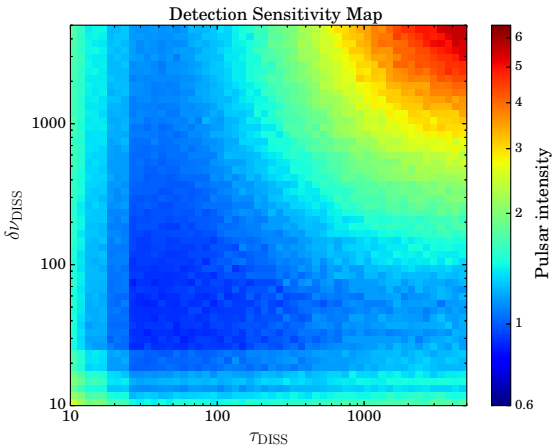}
\caption{Same as Fig.~\ref{sensitivity} but with $N_{\rm{t}}=20$ and $N_{\rm{f}}=20$.}
\label{sensitivity400}
\end{figure}

\begin{figure}
\center
\includegraphics[width=3in]{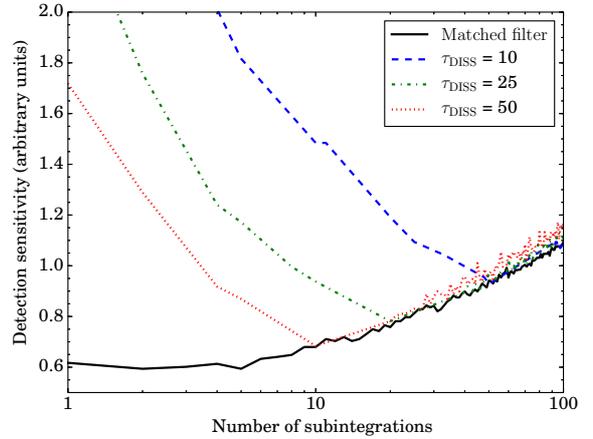}
\caption{Detection sensitivity as a function of number of subintegrations for a given total
integration time. Parameters of the simulation are the same as those of Fig.~\ref{sensitivity}. 
For the solid line, we set $\tau_{\rm{DISS}}=\delta t/2$ and the subintegration time varies 
as $\delta t=T/N_{\rm{t}}$. We also set $\delta\nu_{\rm{DISS}}=\delta\nu/2=50$. For the dashed, 
dash-dotted and dotted line, we set $\delta\nu_{\rm{DISS}}$=10, 25 and 50, respectively.}
\label{npixel}
\end{figure}

With the detection defined above, we carry out simulations to investigate the sensitivity of 
a given survey to pulsars with different scintillation bandwidth and time-scale. We assume that 
the survey has $T=1000$, $B=1000$ and $\sigma_{\rm{n}}=0.1$. In Fig.~\ref{sensitivity}, we 
set $N_{\rm{f}}=10$ and $N_{\rm{t}}=10$, while in Fig.~\ref{sensitivity400} we set $N_{\rm{f}}=20$ 
and $N_{\rm{t}}=20$.
The colour scale of two figures are the same. Both figures show that variance 
images are sensitive to pulsars whose $\delta\nu_{\rm{DISS}}$ and $\tau_{\rm{DISS}}$ are 
close to $\delta\nu/2$ and $\delta t/2$, and the sensitivity drops quickly as $\delta\nu_{\rm{DISS}}$ 
and $\tau_{\rm{DISS}}$ get much smaller or larger than $\delta\nu$ and $\delta t$.
Comparing Fig.~\ref{sensitivity} with Fig.~\ref{sensitivity400}, we can see that for a given total 
bandwidth $B$ and integration time $T$, when we increase the number of channels and subintegrations
we become relatively more sensitive to pulsars with small $\delta\nu_{\rm{DISS}}$ and $\tau_{\rm{DISS}}$, 
and lose sensitivity to pulsars with large $\delta\nu_{\rm{DISS}}$ and $\tau_{\rm{DISS}}$. 
With the same simulation, we can also determine the sensitivity of a Stokes I image with  
five per cent false alarm probability and 80 per cent detection probability. We obtained 
a sensitivity of Stokes I image of $\sim0.25$ for the same simulation parameters, which 
is independent of scintillation time-scale and bandwidth and time and frequency resolutions.

However, for a given total bandwidth and integration time, as we increase the number of 
channels and subintegrations the noise level in the dynamic spectrum  
increases and we lose sensitivity as discussed in Section~\ref{detectionProb}. 
In Fig.~\ref{npixel}, we show the detection sensitivities of the ``matched filter'' case as 
a function of the number of subintegrations for a given total integration time. 
Parameters of simulations are the same as Fig.~\ref{sensitivity}.
We set $\tau_{\rm{DISS}}=\delta t/2$ and the subintegration time varies as $\delta t=T/N_{\rm{t}}$.
We also set $\delta\nu_{\rm{DISS}}=\delta\nu/2=50$ and other parameters same as those 
of Fig.~\ref{sensitivity}. The solid line represents the ``matched filter'' case, and 
in comparison we also present the detection sensitivity of cases that have fixed scintillation 
time-scales with dashed, dash-dotted and dotted lines. 
For the ``matched filter'' case, which gives us the highest sensitivity for different numbers 
of subintegrations, we can see that the sensitivity decreases as the number of subintegrations 
increases. For other cases, the highest sensitivities appear when $\tau_{\rm{DISS}}\approx\delta t/2$, 
which is consistent with Fig.~\ref{matched}.     

\begin{figure*}
\subfloat[]{\includegraphics[width=2in]{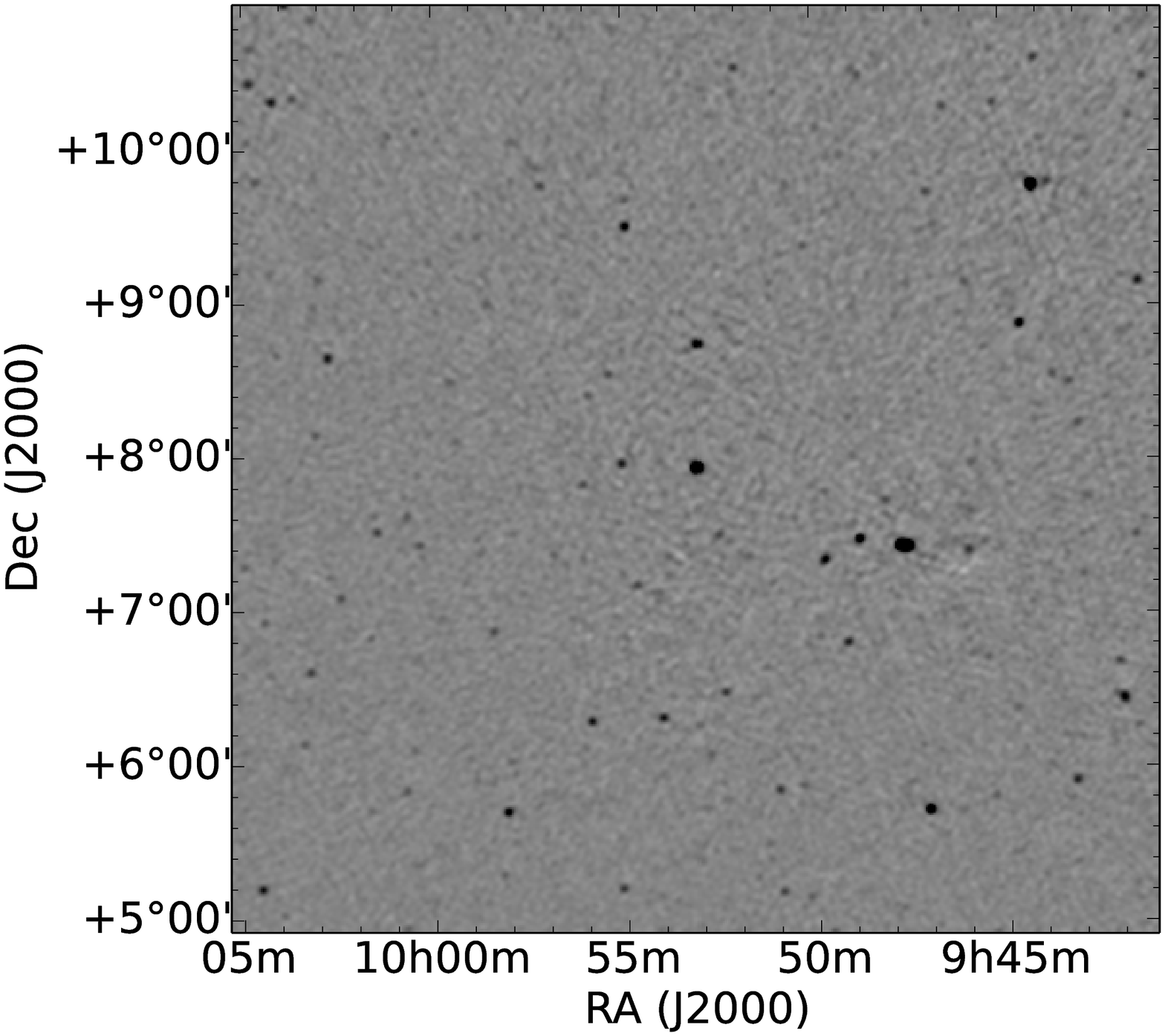}}
\subfloat[]{\includegraphics[width=2in]{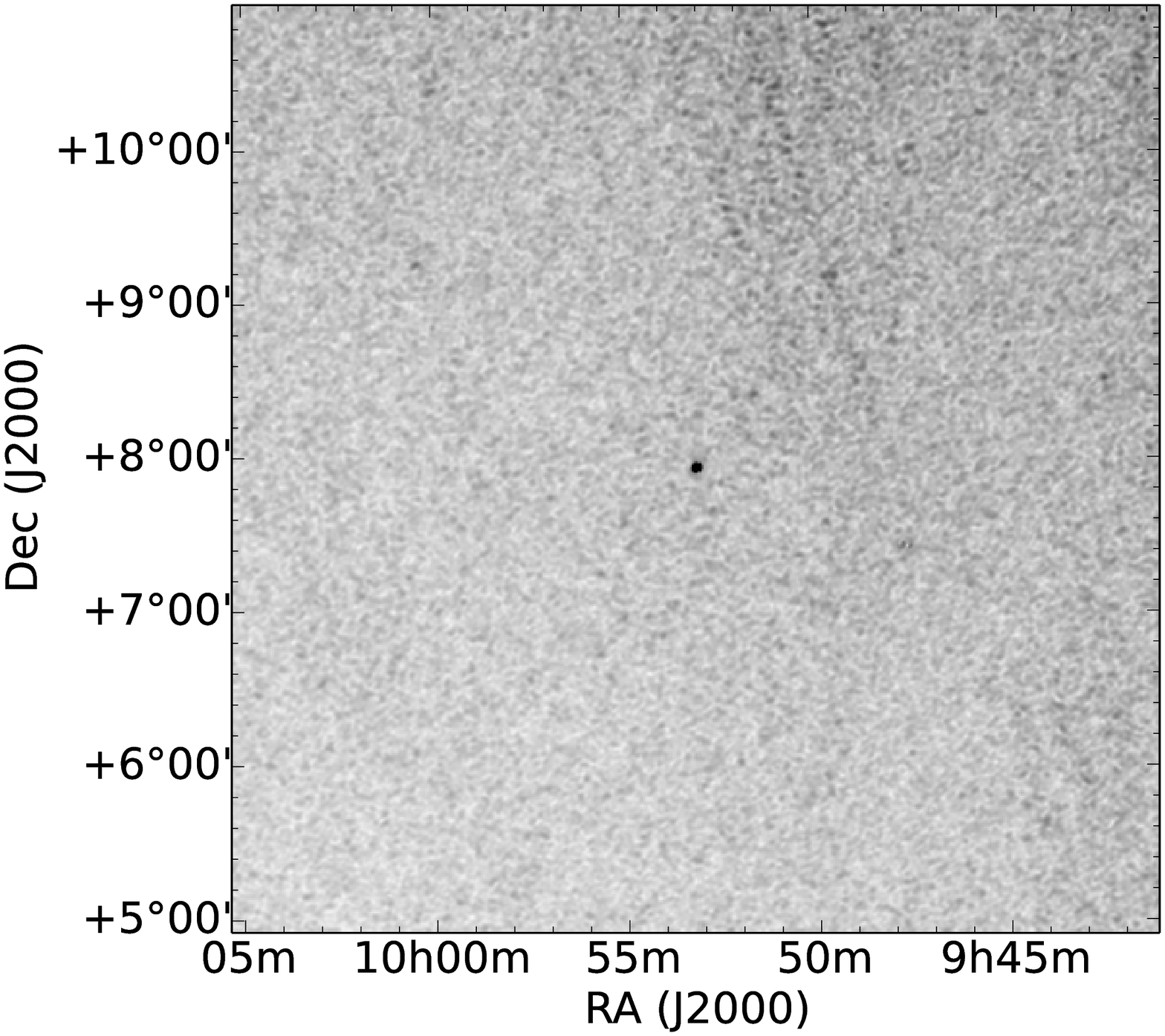}}
\subfloat[]{\includegraphics[width=1.88in]{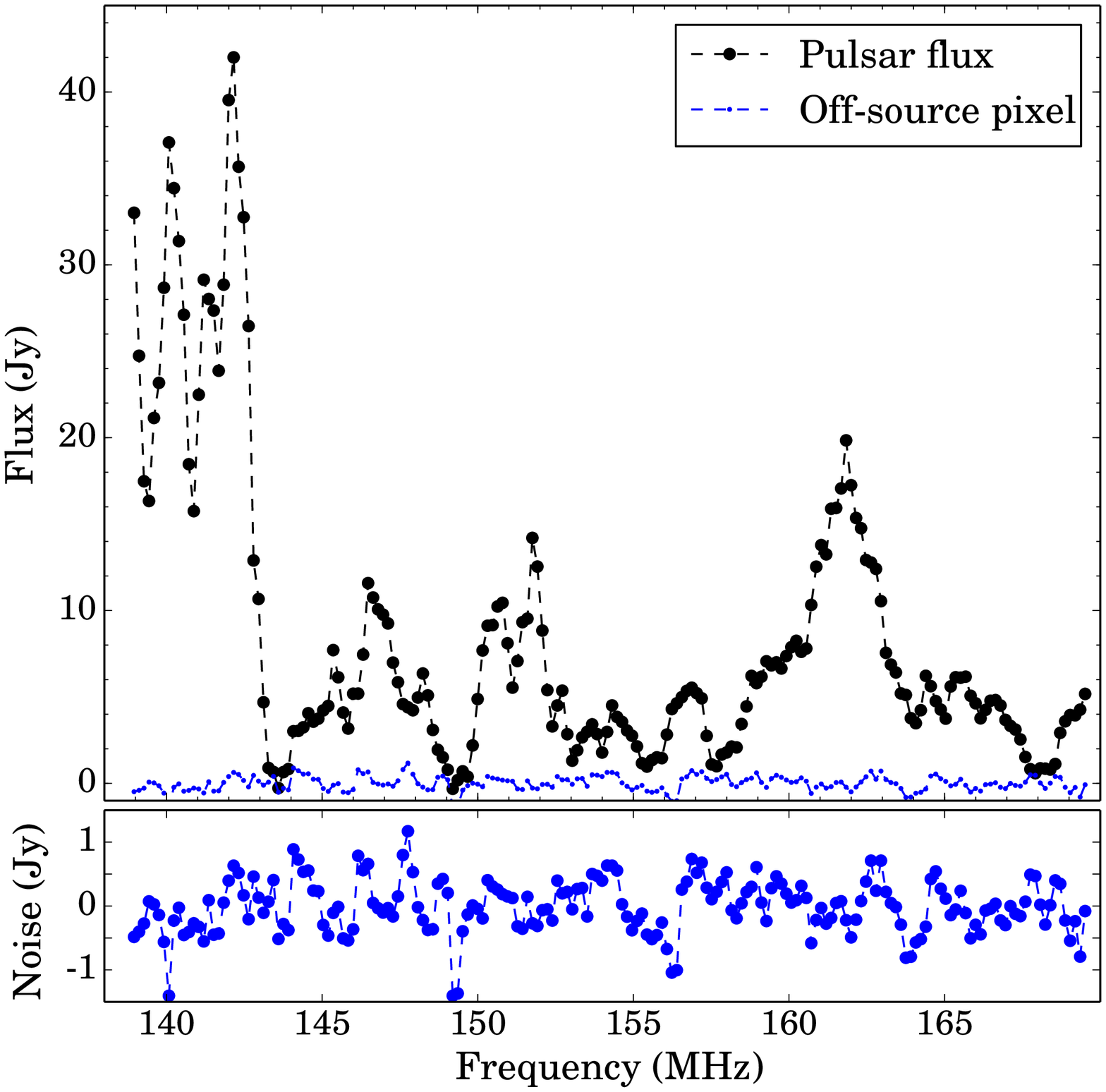}}
\caption{Demonstration of the variance imaging technique. (a) shows the Stokes I image with 
PSR J0953$+$0755 in the centre. (b) shows the variance image of the same field. (c) shows the 
flux densities of PSR J0953$+$0755 and an off-source image pixel as a function of frequency averaged over 
the 112 second snapshot in the upper panel. The bottom panel of (c) shows a zoom-in of flux densities 
of the off-source pixel. For details of the data set see \citet{bmj+16}.}
\label{0953}
\end{figure*}

\section{Demonstration of the technique}
\label{demo}

To demonstrate that we can detect scintillating pulsars in variance images and distinguish them 
from other radio sources, we use data taken with MWA~\citep{tgb+13}. 
For the pulsar PSR J0953$+$0755, \citet{bmj+16} report variability in a time series of 
images taken over a period of approximately 30 minutes. The data were taken at a central frequency of 
154\,MHz with a bandwidth of 30.72\,MHz. For a full discussion of the pulsar variability survey and 
associated data acquisition and reduction methodology see \citet{bmj+16}. Here we examined a 
single 112 second MWA snapshot of PSR J0953$+$0755 whilst undergoing diffractive scintillation. 
The data were re-imaged at 1\,MHz spectral resolution. 
\citet{bmj+16} measure a scintillation bandwidth $\delta \nu_{\rm{DISS}}$ of 4.1\,MHz and 
scintillation time-scale $\tau_{\rm{DISS}}$ of 28.8 minutes. The scintillation pattern in time is 
therefore not resolved, but the 1\,MHz channels are adequate to sample significant frequency fluctuations 
with width of order 4\,MHz.

Fig.~\ref{0953} shows on the left panel a Stokes I image made using the full 30.72\,MHz bandwidth. The 
middle panel shows the variance image made with 1\,MHz channel bandwidth. To make the variance image, we 
have applied a sub-band subtraction filter, where by adjacent channels are subtracted. Such a filter 
removes signals that show slow and smooth fluctuations as a function of frequency, including signals 
from radio sources with non-flat spectra and instrumental effects associated with the bandpass and 
so on. This filter also reduces the S/N of pulsars in the variance image, especially when the 
scintillation is over-sampled in frequency, since fluctuations between channels are reduced when 
the channel bandwidth is narrower than the scintillation bandwidth. 
In the Stokes I image, we can see a number of strong point-like radio sources, which are difficult 
to be distinguished from pulsars without additional information. In contrast, in the variance image PSR 
J0953$+$0755 can be clearly identified and is the only source in the image. 

\begin{figure}
\center
\includegraphics[width=3in]{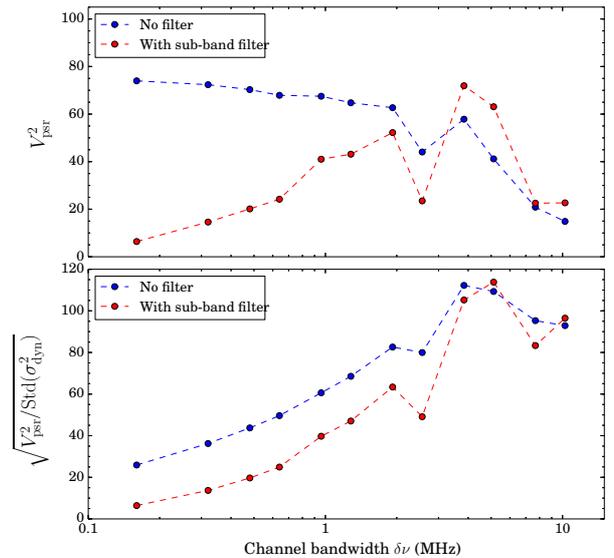}
\caption{Upper panel: $V_{\rm{psr}}$ as a function of channel bandwidth. 
Bottom panel: $\sqrt{V_{\rm{psr}}/\rm{Std}(\sigma_{\rm{dyn}}^{2})}$ as a function of channel bandwidth.
Blue points show results without sub-band subtraction while red points are results after sub-band 
subtraction. }
\label{matchedFilterMWA}
\end{figure}

The right panel of Fig.~\ref{0953} shows the flux density of PSR J0953$+$0755 as a function of 
frequency averaged over the 112 second snapshot. In comparison, we also show the flux density of 
an off-source image pixel as a function of frequency in the upper panel and its zoom-in in the 
bottom panel. The flux densities are measured from images without 
any deconvolution and cleaning, and the frequency resolution of $\sim$160\,kHz is much smaller 
than the scintillation bandwidth. We clearly see the modulation caused by DISS and structures with 
frequency scale of $\sim4$\,MHz. A full dynamic spectrum over 30 minutes has been presented in \citet{bmj+16}. 
In Fig.~\ref{matchedFilterMWA}, the upper and bottom panels show $V_{\rm{psr}}$ and 
$\sqrt{V_{\rm{psr}}/\rm{Std}(\sigma_{\rm{dyn}}^{2})}$ as a function of channel bandwidth, respectively. 
Blue points show results without sub-band subtraction while red points are results after sub-band 
subtraction. 
Without sub-band subtraction, as the channel bandwidth decreases we over-sample the scintillation 
and the variance saturates. After sub-band subtraction the variance decreases as the channel bandwidth 
decreases, and it peaks at around the scintillation bandwidth. Since we cannot resolve the scintillation 
in time and the observing bandwidth is only $\sim7$ times of the scintillation bandwidth, we see 
significant fluctuations in $V_{\rm{psr}}$ as we average the flux over different channel bandwidth. 
To estimate $\sqrt{V_{\rm{psr}}/\rm{Std}(\sigma_{\rm{dyn}}^{2})}$, we assume that the noise is 
Gaussian and $\rm{Std}(\sigma_{\rm{dyn}}^{2})=\sigma_{\rm{n}}^{2}\cdot\sqrt{2N}$. 
$\sigma_{\rm{n}}$, as the standard deviation of noise in the Stokes I image averaged over 30.72\,MHz 
bandwidth, is measured to be $\sim0.075$\,Jy. $\sqrt{V_{\rm{psr}}/\rm{Std}(\sigma_{\rm{dyn}}^{2})}$ 
for both with and without sub-band subtraction cases peak at around a channel bandwidth of 
$\sim5$\,MHz, close to the ``matched filter'' of $\delta\nu/2=\delta\nu_{\rm{DISS}}$ as we discussed in 
Section~\ref{matchedfilter} and \ref{detectionProb}.   

We have assumed that the noise in the image is radiometer noise and is stationary over the image. 
This is a reasonable assumption at centimetre wavelengths but not at meter wavelengths. For MWA the 
system temperature is dominated by the Galactic background and is not stationary over the field of 
view of the image. Incomplete correction of sidelobes of strong sources will not scintillate in time 
but they might show some variation in frequency which could be interpreted as scintillation.
We have used PSR~J0953$+$0755 as an example to show the feasibility of detecting pulsars with 
significant DISS in variance images. Studies of the noise properties and imaging technique
are beyond the scope of this work, though they are of great importance for applying variance 
images in pulsar searching.

\section{Discussion and Conclusions}
\label{discussion}

We investigated the variation of pulsar intensity caused by DISS in variance images. Through simulations, we 
studied the sensitivity of variance images on detecting pulsars and compared it with Stokes I images.
Using data taken with MWA, we demonstrated that variance images can lead to the detection of pulsars and 
distinguish pulsars from other radio sources.
We conclude that DISS of pulsars provides us with a unique way to distinguish pulsars from other radio sources. 
With the variance imaging technique, we will be able to select the most promising pulsar candidates 
from large-scale continuum surveys and enhance the efficiency of following targeted search. 

Variance images are most sensitive to pulsars whose scintillation bandwidth and time-scales are 
close to the channel bandwidth and subintegration time. Therefore, for a given continuum 
survey with certain total bandwidth and integration time, in order to achieve the highest sensitivity 
and detect as many pulsars as possible, we will need to retain frequency and time resolution as high as 
possible, and construct a set of variance images with different channel bandwidth and subintegration time. 
Typically the time scale is not a problem but it is common for the channel bandwidth to limit the detection 
of scintillation.
On the other hand, increasing time and frequency resolution will decrease the overall sensitivity 
of variance images because the noise level in each channel and subintegration increases. This indicates 
that, for a given total bandwidth and integration time, variance images will be relatively less sensitive 
to pulsars with small scintillation bandwidth and time-scales.

The sensitivity maps presented in the paper and simulations we developed can be used to predict 
the number of pulsars detectable with variance images for future large-scale continuum surveys, e.g., 
MWATS, EMU and SKA. Taking the sensitivity of EMU ($\sim10$\,$\mu$Jy) as an example, assuming we have 
enough bandwidth and time and frequency resolution to detect pulsar scintillation, the sensitivity of 
variance images constructed with EMU will be $\sim60$ to 100\,$\mu$Jy, depending on the time and 
frequency resolution. To determine the number of pulsars that can be detected, pulsar Galactic and 
flux density distributions and their scintillation time-scales and bandwidths will have to be considered.  
We defer studies of pulsar population and prediction of pulsar detection with variance images to 
future work.

Although variance images allow unique identifications of pulsars, for given false alarm and detection 
probabilities, the sensitivity of variance images is lower than that of Stokes I images. 
Therefore, very faint pulsars detectable in continuum surveys might not be identified in variance images
even if they show strong scintillation. However, the diffractive scintillation features of pulsars 
can still be powerful criteria to distinguish them from other radio sources. Instead of making 
variance images, we could first identify point sources in continuum surveys and then use scintillation 
features to distinguish pulsars from other sources.

\section*{Acknowledgements}

This scientific work makes use of the Murchison Radioastronomy
Observatory, operated by CSIRO. We acknowledge
the Wajarri Yamatji people as the traditional owners
of the Observatory site. Support for the operation of the
MWA is provided by the Australian Government Department
of Industry and Science and Department of Education
(National Collaborative Research Infrastructure Strategy:
NCRIS), under a contract to Curtin University administered
by Astronomy Australia Limited.  The Parkes radio telescope is part 
of the Australia Telescope, which is funded by the Commonwealth of 
Australia for operation as a National Facility managed by the 
Commonwealth Scientific and Industrial Research Organisation (CSIRO)

%%%%%%%%%%%%%%%%%%%%%%%%%%%%%%%%%%%%%%%%%%%%%%%%%%

\bibliography{ms}
%%%%%%%%%%%%%%%%%%%%%%%%%%%%%%%%%%%%%%%%%%%%%%%%%%

% Don't change these lines
\bsp	% typesetting comment
\label{lastpage}
\end{document}